# An Improved Structure Of Reversible Adder And Subtractor

Aakash Gupta [1], Pradeep Singla [2], Jitendra Gupta [3], Nitin Maheshwari [4]

[1]*Deptt. Of Computer Science,* [2,3,4] *Department of Electronics and Communication Engineering,*
*Sonipat Institute of Engineering & Mgmt.*

[1]*aakashgarg1987@gmail.com*
[2]*pradeepsingla7@gmail.com*
[3]*gupta.92.jitendra@gmail.com*
[4]*nitinmaheshwari62@yahoo.com*

**Abstract-** In today's world everyday a new technology which is faster, smaller and more complex than its predecessor is being developed. The increased number of transistors packed onto a chip of a conventional system results in increased power consumption that is why Reversible logic has drawn attention of Researchers due to its less heat dissipating characteristics. Reversible logic can be imposed over applications such as quantum computing, optical computing, quantum dot cellular automata, low power VLSI circuits, DNA computing. This paper presents the reversible combinational circuit of adder, subtractor and parity preserving subtractor. The suggested circuit in this paper are designed using Feynman, Double Feynman and MUX gates which are better than the existing one in literature in terms of Quantum cost, Garbage output and Total logical calculations.

**Keywords –**Reversible Logic, Constant Input, Garbage output, Total Logical Calculation, Adder and Subtractor.

## I. INTRODUCTION

In electronics hardware designing energy dissipation is one of the most important aspects. The concept of reversibility in digital circuits is firstly related to energy by Landauer in 1961 who stated that there is small amount of heat dissipation the circuit due to loss of one bit of information and it would be equal to kTln2 where 'k' is Boltzman constant and T is the temperature[1] . Also in 1973 it was proved by Bennett that the energy kTln2 would not be dissipate from the circuit if input can be extracted from output and it would be possible if and only if reversible gates are used[2]. A circuit will be reversible if input vector can be specifically retrieved from output vectors and here is one to one correspondence between input and output [3]. Thermodynamics explain the concept of reversibility which taught the benefits of reversibility over irreversibility. Reversible logic synthesis of reversible combinational logic differs from sequential logic in that the output of the logic device depends on the present input unlike sequential circuits in which output depends on present as well as past input too.

In this paper combinational circuits have been synthesized using reversible gates and this paper provides the modified design of adder and subtractor in terms of garbage output, total number of calculations, quantum cost. Double Feynman, Feynman and MUX gates have been used to design the circuits.

This paper is organised in the following prospective: Section ll describe the Grassroots of the reversible logic and the conditions for the reversibility. Different reversible gate, their structure and Quantum cost is also discussed in the same section. In Section III, We have discussed About Combinational circuits of Adder and subtractor . In section IV, the proposed design of reversible Adder/ subtactor and parity preserving subtractor is described. In the Section V, the numerical results of proposed reversible adder/ subtractor and parity preserving subtractor is shown.

## II. PROPOSED ALGORITHM GRASSROOTS OF REVERSIBLE LOGIC

In reversible logic there should be same number of input and outputs i.e., the number of input lines and output lines must be same in a circuit and there must be one to one mapping between input and output. The gate must be running in both directions i.e., the input can be retrieved from the output, when a digital circuit obeys these two stipulations then the second law of thermodynamics guarantees that no heat dissipation occur in the circuit[4]. In the synthesis of reversible circuit's direct fan out and feedback is not permitted. A reversible circuit should be designed using minimum number of reversible logic gates in order to achieve efficiency and less complexity. There are some parameter in reversible circuit design for determining the complexity and performance of circuit, such as garbage output, quantum cost, constant inputs and total logical calculation [5]. Some important terms used in reversible logics are Garbage Output refers to the number of unused outputs present in a reversible logic circuit.

**Quantum cost** [3] refers to the cost of circuits in terms of the cost of primitive gates. It is calculated knowing the number primitive reversible logic gates (1*1 or 2*2) required to realize the circuit. **Total Logical calculation [6]** is the count of the XOR, AND, NOT logic in the output circuit. **Constant inputs [7]** are the number of inputs that are to be maintained constant at either 0 or 1 in order to synthesize the given logical function.

A. *Radical Reversible Logic Gates –*

  *1) Feynman Gate [3] [10]:*

Fig.1 shows the pictorial representation of 3×3 reversible Feynman gate [10]. It has three inputs (A, B) and three outputs (P, Q). The outputs are defined by P=A, Q=A Xor B. Quantum cost of a Feynman gate is 1.

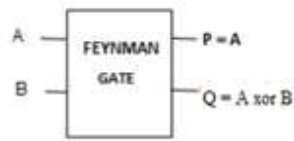

Fig. 1 Feynman gate

  *2) Double Feynman Gate (F2G [8]):*

Fig.2 shows the pictorial representation of 3×3 and R= A Xor C. Quantum cost of a Double Feynman gate is 2. It has three inputs (A, B, C) and three outputs (P, Q, R). The outputs are defined by P=A, Q=A Xor B, R= A xor C.

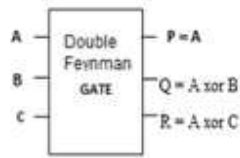

Fig. 2 Double Feynman gate

  *3) Fredkin Gate [9] [11]:*

Fig.3 shows the pictorial representation of 3×3 reversible Fredkin gate [9]. It has three inputs (A, B, C) and three outputs (P, Q, R). The outputs are defined by P=A, Q=A'B Xor AC and R= A'C XOR AB. Quantum cost of a Fredkin gate is 5.

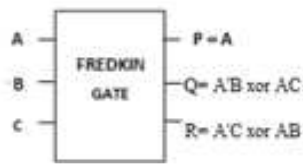

Fig.3 Fredkin gate

  *4) MUX Gate [3]:*

**An Improved Structure Of Reversible Adder And Subtractor**

Fig.4 shows the pictorial representation of 3×3 reversible MUX (MG) gate [3]. It has three inputs (A, B, C) and three outputs (P, Q, R). The outputs are defined by P=A, Q=A Xor B Xor C and R= A'C Xor AB. Quantum cost of a MUX gate is 4.

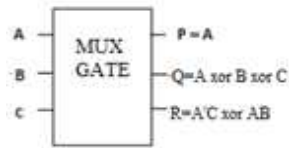

Fig.4 MUX gate

### III. COMBINATIONAL CIRCUIT

Combinational circuits consist of logic gates where output at any instant are determined only by present combination of inputs without regard of previous inputs or previous state of the output. Some commonly used combinational circuits are adder, subtractor etc

*A. Adders-*

Adder is one of the most basic combinational circuits that perform addition of the bits. Addition of the binary digits is the most basic arithmetic operation[12]. Adders are classified into two possible combinations [13]

  *1) Half adder:*

The circuit that performs the simple addition of two bits is known as half adder in the fig.5 two inputs (A, B) and two outputs (S, C) are present

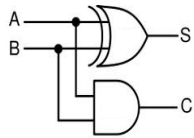

Fig 5. Irreversible Half Adder

  *2) Full adder:*

Full adder performs addition of three bits i.e. A,B and Cin and output Sum(S) and Carry(C). A full adder can add the same two input bits as a full adder PLUS an extra bit for an incoming carry as shown in fig.6. This is important for cascading adders together to create N-bit adders.

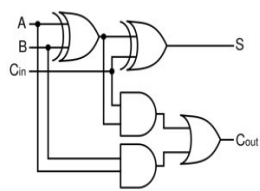

Fig.6. Irreversible full adder

*B. Subtractor:*

A Combinational circuit which goes on performing subtraction of bits is known as subtractor. Subtractor are further classified into 2 parts

  *1) Half Subtractor:*

Subtraction of two bits takes place in the half subtractor and two outputs are produced i.e. difference(D) and borrow(B).Figure 7 shows the systematic figure of half subtractor in which X and Y are input bits and D and B are output bits

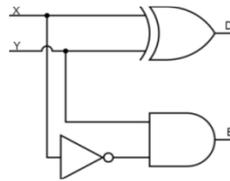

Fig.7. Irreversible half subtractor

*2) Full Subtractor:*

In full subtractor, subtraction of three bit is carried out i.e A, B, Bin and output difference and borrow(Bout) is produced. Fig 8 below shows the systematic figure of full subtractor, in which two EXOR, two AND, two NOT and one OR gate is used in a specific combination.

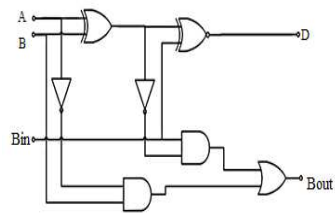

Fig.8. Irreversible full subtractor

IV. PROPOSED ADDER AND SUBTRACTOR

Feynman gate works in copy mode and complement mode. In above circuit the Feynman gate is implied in copy mode. Both inputs have been fed to separate Feynman gate and carry in input to a separate Feynman gate FG3 .the output of all above three Feynman gates is being fed to FG4 whose output line is SUM or difference line. For the carry out MUX gates have been used by providing sum of A and B as one input to MUX gate. Fig 9.shows reversible half adder/subtractor and Fig10. Shows Reversible full adder/subtractor.

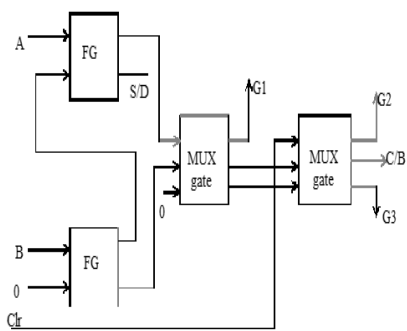 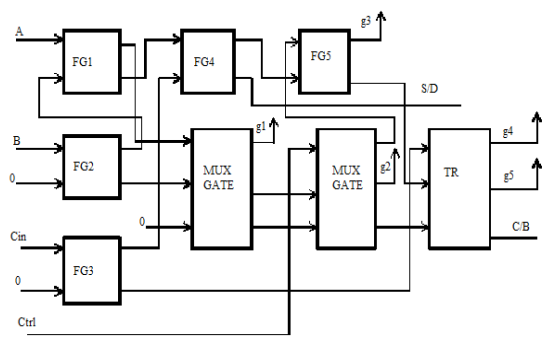

Fig.9. reversible half Adder/subtractor        Fig.10 reversible half Adder/subtractor

**An Improved Structure Of Reversible Adder And Subtractor**

*A. Parity preserving reversible half-subtractor circuit-*

Half-subtractor circuit executes subtraction operation if A and B be inputs then, the output equations of the Borrow and difference are as follows:

$$Diff = A \text{ xor } B, Borrow = A'B$$

The proposed parity preserving reversible half-subtractor circuit is shown in Fig11.this circuit is composed of F2G and MUX gates . the quantum cost of F2G is 2 and MUX is 4. so, the quantum cost of this circuit is 6. The proposed circuit requires two constant inputs and produces two garbage outputs. Proposed parity preserving reversible half-subtractor circuit can be used in constructing fault tolerant reversible circuits in which there would be no necessity of parity bit for error detection.

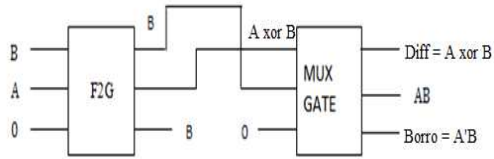

Fig. 11 Proposed parity preserving reversible half-subtractor circuit

*B. Parity preserving reversible full-subtractor circuit-*

Full subtractor circuit executes A-B-C operation if A, B and C be inputs so, the equations of borrow and difference is as follows:

$$Diff = A \text{ xor } B \text{ xor } C, Borr = A'B \text{ xor } A'C \text{ xor } BC$$

The proposed parity preserving reversible full-subtractor circuit is shown in fig12.this circuit is composed of 3 F2G gate and 1 MUX gate. It produces 4 garbage outputs and requires only one constant input. The quantum cost of F2G is 2 AND MUX is 4. So, the Quantum cost of this circuit is 10. proposed parity preserving reversible full-subtractor circuit can be used for designing fault tolerant reversible systems which is the necessary requirement of nanotechnology based systems.

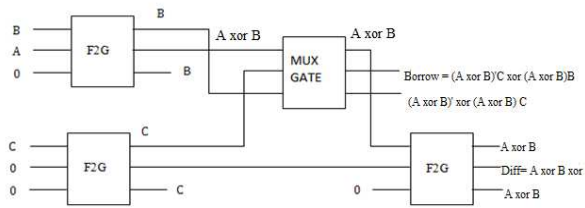

Fig.12 Proposed parity preserving reversible full-subtractor circuit

## V. NUMERAL RESULT

*A. Quantum cost-*

*1) Quantum cost of half adder/ subtractor:*

In the reversible half adder/subtractor 2 Mux gates and 2 Feynman gates is being used for the proposed designs Now let 'm' is the Quantum cost of Mux gate and 'F' is the Quantum cost of the Feynman gate. So, the Quantum cost (A) of the Reversible half adder/subtractor is

$$\text{Quantum cost (A/S)} = 2m+2F$$

*2) Quantum cost of Full adder/ subtractor:*

Similarly in reversible Full adder/subtractor , 2 Mux gate, 1 TR Gate and 5 Feynman gate has been used .

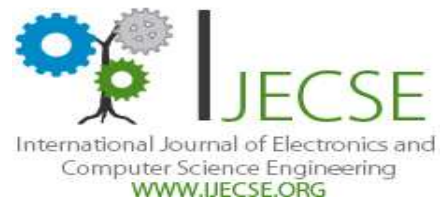

So, Now let 'm' is the Quantum cost of Mux gate, TR is the Quantum cost of TR gate and 'F' is the Quantum cost of the Feynman gate. so, the Quantum cost(A) of the Reversible full adder/subtractor is

$$\text{Quantum cost}(A/S) = 2m+5F+1TR$$

*3) Quantum cost of 3 bit reversible parity preserving half subtractor:*

In reversible parity preserving half subtractor we have used 1 mux gate and 1 Double Feynman gate.
Now let 'm' be the Quantum cost of the mux gate and 'D' is the Quantum cost of Double Feynman gate . so, the Quantum cost of reversible parity preserving half subtractor is

$$\text{Quantum cost (HA)} = 1m+1D$$

*4) Quantum cost of 3 bit reversible parity preserving full-subtractor :*

In reversible parity preserving full-subtractor we have used 1 Mux gate and 3 Double Feynman gate.
Now let m is the Quantum cost of MUX gate and D is the Quantum cost of Double Feynman gate. So, the Quantum cost of reversible parity preserving full-subtractor is

$$\text{Quantum cost (FA)} = 1m+3D$$

**B. Total Logical calculation (T):**

Assuming
$\alpha$ = A two input XOR gate calculation
$\beta$ = A two input AND gate calculation
$\delta$ = A NOT gate calculation
T = Total logical calculation

The Total logical calculation is the count of the XOR, AND, NOT logic in the output circuit. For example MUX gate has three XOR gate and two AND gate and one NOT gate in the output expression. Therefore $(M) = 3\alpha+2\beta+\delta$.and Double Feynman gate has two XOR gate .Therefore T (D) =2 $\alpha$

*1) Total Logical calculation (T) of reversible half adder/ subtractor:*

In reversible half adder/ subtractor we have used 2 Mux gate and 2 Feynman gates. So Total logical calculation of reversible half adder/ subtractor is

$$T = 2\times (3\alpha+2\beta+\delta)(\text{ for MUX gate}) + 2\times\alpha(\text{for Feynman gate})$$
$$T(A/S) = 8\alpha+6\beta+2\delta$$

*2) Total Logical calculation (T) of reversible Full adder/ subtractor:*

In reversible Full adder/ subtractor we have used 2 Mux gate,1 TR gate and 5 Feynman gates. So Total logical calculation of reversible half adder/ subtractor is

$$T = 2\times(3\alpha+2\beta+\delta)(\text{ for MUX gate}) + 5\times1\alpha(\text{for Feynman gate}) + 1(2\alpha+\beta+\delta) \text{ (for TR gate)}$$
$$T(A/S) = 13\alpha +5\beta +3\delta$$

*3) Total Logical calculation (T) of reversible parity preserving half subtractor:*

In reversible parity preserving half subtractor we have used 1 mux gate and 1 Double Feynman gate. So, Total Logical calculation (T) of reversible parity preserving half subtractor

$$T(HA) = 1\times(3\alpha+2\beta+\delta)(\text{for MUX gate}) + 1\times2\alpha(\text{for Double Feynman gate})$$
$$T = 5\alpha +2\beta +\delta$$

*4) Total Logical calculation (T) of reversible parity preserving full-subtractor :*

In reversible parity preserving full-subtractor we have used 1 MUX gate and 3 Double Feynman gate .So,Total Logical calculation (T) of reversible parity preserving full-subtractor

$$T(FA) = 1\times(3\alpha+2\beta+\delta) \text{ ( for MUX gate)} + 3\times2\alpha(\text{for Double Feynman gate})$$
$$T = 9\alpha +2\beta +\delta$$

Now from the discussion of the numerical results of the proposed work it is clear that the proposed design is better than the existing one as the Quantum cost of the Existing half adder/ subtractor and full adder /subtractor is 2f+2fr and 5f+2fr+1TR respectively .and the Quantum cost of proposed design of half adder/subtractor and full adder/ subtractor 2m+2F and 2m+5F+1TR respectively. The Quantum cost of existing parity preserving half subtractor and

**An Improved Structure Of Reversible Adder And Subtractor**

full subtractor is 1fr+1D and 1fr+3D respectively. And the Quantum cost of proposed parity preserving half subtractor and full subtractor is 1m+1D and 1m+3D respectively. Similarly in the previous work the total logical calculation of half adder/ subtractor and full adder/subtractor is $6\alpha +8\beta +4\delta$ and $10\alpha +9\beta +4\delta$ respectively. And the total logical calculation of the proposed design of half adder/subtractor and full adder/subtractor is $8\alpha +6\beta +2\delta$ and $13\alpha+5\beta+3\delta$ respectively and the total logical calculation of existing reversible parity preserving half subtractor and full subtractor is $4\alpha +4\beta +2\delta$ and $10\alpha +4\beta +2\delta$ respectively. And the total logical calculation of the proposed design of reversible parity preserving half subtractor and full subtractor is $5\alpha +2\beta +\delta$ and $9\alpha +2\beta +\delta$ respectively

## VI. CONCLUSIONS

The Adder and subtractor are mostly used for applications like Arithematic and logical unit(ALU), Programme status word (PSW), Calculators, Embedded system, seven segment display etc.

This paper suggested a new improved design of Adder, Subtractor and Parity preserving Subtractor with the help of Feynman gate, Double Feynman gate and MUX gate. Since the mux gate has lower quantum cost as compared to the Fredkin gate defined in literature and mux gate can all the operation as do as Fredkin. The numeral results are also shown in the paper which shows the optimized results of the proposed design of combinational circuits against the previous ones and will provides a new arena to design digital logical systems.